\documentclass{aa}  
\usepackage{txfonts}
\usepackage{natbib}
\bibpunct{(}{)}{;}{a}{}{,}
\usepackage{graphicx}
\begin{document}
   \title{The HARPS search for southern extra-solar planets.}

   \subtitle{IX. Exoplanets orbiting HD\,100777, HD\,190647, and HD\,221287\thanks{Based on 
   observations made with the HARPS instrument on the ESO 3.6-m telescope at the La Silla Observatory (Chile) under the GTO 
   programme ID 072.C-0488.}}

   \author{
D.~Naef \inst{1,2}
          \and
	  M. Mayor \inst{2}
	  \and
	  W.~Benz \inst{3}
	  \and
	  F.~Bouchy \inst{4}
	  \and
	  G.~Lo~Curto \inst{1}
	  \and
	  C.~Lovis \inst{2}
	  \and
	  C.~Moutou \inst{5}
	  \and
	  F.~Pepe \inst{2}
	  \and
	  D.~Queloz \inst{2}
	  \and
	  N.C.~Santos \inst{2,6,7}
	  \and
	  S.~Udry \inst{2}
          }

   \offprints{D.~Naef \email{dnaef@eso.org}}

   \institute{
European Southern Observatory, Casilla 19001, Santiago 19, Chile
             \and
             Observatoire Astronomique de l'Universit\'e de Gen\`{e}ve, 51 Ch. des Maillettes, CH-1290 Sauverny, Switzerland 
	     \and
	     Physikalisches Institut Universit\"{a}t Bern, Sidlerstrasse 5, CH-3012 Bern, Switzerland
	     \and
	     Institut d'Astrophysique de Paris, UMR7095, Universit\'{e} Pierre \& Marie Curie, 98 bis Bd Arago, F-75014 Paris, France
	     \and
     	     Laboratoire d'Astrophysique de Marseille, Traverse du Siphon, F-13376 Marseille 12, France
	     \and
	     Centro de Astronomia e Astrof\'{i}sica da Universidade de Lisboa, Observat\'{o}rio Astron\'{o}mico de 
	     Lisboa, Tapada da Ajuda, P-1349-018 Lisboa, Portugal
             \and
	     Centro de Geof\'{i}sica de \'{E}vora, Rua Rom\~{a}o Ramalho 59, P-7002-554 \'{E}vora, Portugal
             }

   \date{Received 27 February 2007 / Accepted 11 April 2007}

 
  \abstract
   {
   The {\footnotesize HARPS} high-resolution high-accuracy spectrograph was made available to the astronomical 
   community in the second half of 2003. Since then, we have been using this instrument for monitoring radial velocities 
   of a large sample of Solar-type stars ($\simeq$\,1400 stars) in order to search for their possible low-mass companions.
   }
   { 
   Amongst the goals of our survey, one is to significantly increase the number of detected extra-solar planets in a 
   volume-limited sample  to improve our knowledge of their orbital elements distributions and thus obtain better 
   constraints for planet-formation models.
   }
   {Radial-velocities were obtained from high-resolution {\footnotesize HARPS} spectra via the
   cross-correlation method. We then searched for Keplerian signals in the obtained radial-velocity data sets. 
   Finally, companions orbiting our sample stars were characterised using the fitted orbital parameters.
   }
   {
   In this paper, we present the {\footnotesize HARPS} radial-velocity data and orbital solutions for 3 Solar-type stars: 
   {\footnotesize HD}\,100777, {\footnotesize HD}\,190647, and {\footnotesize HD}\,221287. 
   The radial-velocity data of {\footnotesize HD}\,100777 is best explained by the presence of a 1.16\,M$_{\rm Jup}$ 
   planetary companion on a 384--day eccentric orbit ($e$\,=\,0.36). The orbital fit obtained for the slightly evolved star 
   {\footnotesize HD}\,190647 reveals the presence of a long-period ($P$\,=\,1038\,d) 1.9\,M$_{\rm Jup}$ planetary 
   companion on a moderately eccentric orbit ($e$\,=\,0.18). {\footnotesize HD}\,221287 is hosting a 3.1\,M$_{\rm Jup}$ 
   planet on a 456--day orbit. The shape of this orbit is not very well-constrained  because of our non-optimal temporal 
   coverage and because of the presence of abnormally large residuals. We find clues that these large residuals result from 
   spectral line-profile variations probably induced by processes related to stellar activity.
   }
   {}

   \titlerunning{The HARPS search for southern extra-solar planets IX}
   \authorrunning{D.~Naef et al.}
   \maketitle
%

\section{Introduction}\label{Intro}
The {\sl High Accuracy Radial-velocity Planet Searcher} 
\citep[{\footnotesize HARPS},][]{Pepemessenger, Pepe330075, Mayormessenger} was put in operation during the second half of 2003. 
{\footnotesize HARPS} is a high-resolution, fiber-fed echelle spectrograph mounted on the 3.6--m telescope at 
{\footnotesize ESO}--La Silla Observatory (Chile). It is placed in a vacuum vessel and is accurately thermally-controlled 
(temperature va\-ria\-tions are less than 1\,mK over one night, less than 2\,mK over one month). Its most striking characteristic is 
its unequaled stability and radial-velocity ({\footnotesize RV}) accuracy: 1\,m\,s$^{\rm -1}$ in routine o\-pe\-ra\-tions. A 
sub-m\,s$^{\rm -1}$ accuracy can even be achieved for inactive, slowly rotating stars when an optimized observing strategy aimed 
at ave\-ra\-ging out the stellar oscillations signal is applied \citep{Santosmuara, Lovis3neptunes}. 

The {\footnotesize HARPS} Consortium that manufactured the instrument for {\footnotesize ESO} has received Guaranteed Time 
Observations ({\footnotesize GTO}). The core programme of the {\footnotesize HARPS-GTO} is a very high 
{\footnotesize RV}-precision search for low-mass planets around non-active and slowly rotating Solar-type stars. Another programme 
carried out by the {\footnotesize HARPS-GTO} is a lower {\footnotesize RV} precision planet search. It is a survey of about 850 
Solar-type stars at a precision better than 3\,m\,s$^{\rm -1}$. The sample is a volume-limited complement (up to 57.5\,pc) of the 
{\footnotesize CORALIE} sample \citep{UdrycoralieII}. The goal of this sub-programme is to obtain improved Jupiter-sized planets 
orbital elements distributions by substantially increasing the size of the exoplanets sample. Statistically robust  orbital 
elements distributions put strong constraints on the various planet formation scenarios. The total number of extra-solar planets 
known so far is over 200. Nevertheless, some sub-categories of planets with special characteristics (e.g. hot-Jupiters or very 
long-period planets) are still weakly populated. The need for additional detections thus remains high.

With typical measurement precisions of 2-3\,m\,s$^{\rm -1}$, the {\footnotesize HARPS} volume-limited programme does not 
necessarily aim at detecting very low-mass planetary companions, and it is mostly sensitive to planets that are more massive than 
Saturn. To date, it has already allowed the detection of several short-period pla\-nets: {\footnotesize HD}\,330075\,b 
\citep{Pepe330075}, {\footnotesize HD}\,2638\,b, {\footnotesize HD}\,27894\,b, 
{\footnotesize HD}\,63454\,b \citep{Moutoushortp}, and {\footnotesize HD}\,212301\,b \citep{Locurto212301}.

In this paper, we report the detection of 3 longer-period pla\-ne\-ta\-ry companions orbiting stars in the volume-limited sample: 
{\footnotesize HD}\,100777\,b, {\footnotesize HD}\,190647\,b, and {\footnotesize HD}\,221287\,b. In 
Sect.~\ref{Stars}, we describe the characteristics of the 3 host stars. In Sect.~\ref{RVdata}, we present our 
{\footnotesize HARPS} radial-velocity data and the orbital solutions for the 3 targets. In Sect.~\ref{oc221287}, we discuss the 
origin of the high residuals to the orbital solution for {\footnotesize HD}\,221287. Finally, we summarize our findings 
in Sect.~\ref{Conc}. 

\begin{table}[t!]
\caption{
\label{tabstars}
Observed and inferred stellar characteristics of HD\,100777, HD\,190647, and HD\,221287 (see text for details).
}
\begin{tabular}{ll|r@{\,$\pm$\,}lr@{\,$\pm$\,}lr@{\,$\pm$\,}l}
\hline\hline
\noalign{\vspace{0.05cm}}
\multicolumn{2}{c|}{} & \multicolumn{2}{c}{HD\,100777} & \multicolumn{2}{c}{HD\,190647} & \multicolumn{2}{c}{HD\,221287} \\
\hline
\noalign{\vspace{0.05cm}}
{\footnotesize HIP} &                        & \multicolumn{2}{c}{56572}                        & \multicolumn{2}{c}{99115}                        & \multicolumn{2}{c}{116084}\\[0.1cm]
Type                &                        & \multicolumn{2}{c}{K0}                           & \multicolumn{2}{c}{G5}                           & \multicolumn{2}{c}{F7V}\\[0.1cm]
$m_{\rm V}$         &                        & \multicolumn{2}{c}{8.42}                         & \multicolumn{2}{c}{7.78}                         & \multicolumn{2}{c}{7.82}\\[0.1cm]
$B-V$               &                        & \multicolumn{2}{c}{0.760}                        & \multicolumn{2}{c}{0.743}                        & \multicolumn{2}{c}{0.513}\\[0.1cm]
$\pi$               & $[$mas$]$              & 18.84 & 1.14                                     & 18.44 & 1.10                                     & 18.91 & 0.82\\[0.1cm]
$d$                 & $[$pc$]$               & \multicolumn{2}{c}{52.8$^{\rm +3.4}_{-\rm 3.0}$} & \multicolumn{2}{c}{54.2$^{\rm +3.4}_{-\rm 3.1}$} & \multicolumn{2}{c}{52.9$^{\rm +2.4}_{-\rm 2.2}$}\\[0.1cm]
$M_{\rm V}$         &                        & \multicolumn{2}{c}{4.807}                        & \multicolumn{2}{c}{4.109}                        & \multicolumn{2}{c}{4.203}\\[0.1cm]
$B.C.$              &                        & \multicolumn{2}{c}{$-$0.119}                     & \multicolumn{2}{c}{$-$0.109}                     & \multicolumn{2}{c}{$-$0.010}\\[0.1cm]
$L$                 & $[$L$_{\odot}$$]$      & \multicolumn{2}{c}{1.05}                         & \multicolumn{2}{c}{1.98}                         & \multicolumn{2}{c}{1.66}\\[0.1cm]
$T_{\rm eff}$       & $[$K$]$                & 5582 & 24                                        & 5628 & 20                                        & 6304 & 45\\[0.1cm]
$\log g$            & $[$cgs$]$              & 4.39 & 0.07                                      & 4.18 & 0.05                                      & 4.43 & 0.16\\[0.1cm]
$[$Fe/H$]$          &                        & 0.27 & 0.03                                      & 0.24 & 0.03                                      & 0.03 & 0.05\\[0.1cm]
$V_{\rm t}$         & $[$km\,s$^{\rm -1}$$]$ & 0.98 & 0.03                                      & 1.06 & 0.02                                      & 1.27 & 0.12\\[0.1cm]
$M_*$               & $[$M$_{\odot}$$]$      & 1.0  & 0.1                                       & 1.1  & 0.1                                       & 1.25 & 0.10\\[0.1cm]
$\log R^{'}_{HK}$   &                        & \multicolumn{2}{c}{$-$5.03}                      & \multicolumn{2}{c}{$-$5.09}                      & \multicolumn{2}{c}{$-$4.59}\\[0.1cm]
$P_{\rm rot}$       & $[$d$]$                & 39 & 2                                           & 39 & 2                                           & 5.0 & 2\\[0.1cm]
$age$               & $[$Gyr$]$              & \multicolumn{2}{c}{$>$2}                         & \multicolumn{2}{c}{$>$2}                         & \multicolumn{2}{c}{1.3}\\[0.1cm]
$v\sin i$           & $[$km\,s$^{\rm -1}$$]$ & 1.8  & 1.0                                       & 2.4  & 1.0                                       & 4.1  & 1.0\\[0.05cm]
\hline
\end{tabular}
\end{table}

\section{Stellar characteristics of HD\,100777, HD\,190647, and HD\,221287}\label{Stars}

\begin{table}[t!]
\caption{
\label{hd100777data}
HARPS radial-velocity data obtained for HD\,100777.
}
\begin{tabular}{ccc}
\hline\hline
\noalign{\vspace{0.05cm}}
Julian date & RV & Uncertainty \\
BJD\,$-$\,2\,400\,000 $[$d$]$ & \multicolumn{2}{c}{$[$km\,s$^{\rm -1}]$}\\
\hline
\noalign{\vspace{0.05cm}}
53\,063.7383 & 1.2019 & 0.0016 \\
53\,377.8740 & 1.2380 & 0.0015 \\
53\,404.7626 & 1.2205 & 0.0014 \\
53\,407.7461 & 1.2164 & 0.0014 \\
53\,408.7419 & 1.2176 & 0.0016 \\
53\,409.7891 & 1.2161 & 0.0015 \\
53\,468.6026 & 1.2109 & 0.0016 \\
53\,470.6603 & 1.2122 & 0.0021 \\
53\,484.6703 & 1.2273 & 0.0013 \\
53\,489.5948 & 1.2319 & 0.0022 \\
53\,512.5671 & 1.2494 & 0.0017 \\
53\,516.5839 & 1.2539 & 0.0014 \\
53\,518.5962 & 1.2554 & 0.0021 \\
53\,520.5778 & 1.2558 & 0.0022 \\
53\,543.5598 & 1.2649 & 0.0013 \\
53\,550.5289 & 1.2641 & 0.0016 \\
53\,573.4689 & 1.2682 & 0.0014 \\
53\,579.4705 & 1.2641 & 0.0022 \\
53\,724.8617 & 1.2520 & 0.0012 \\
53\,762.8169 & 1.2355 & 0.0013 \\
53\,765.7645 & 1.2311 & 0.0016 \\
53\,781.8894 & 1.2242 & 0.0016 \\
53\,789.7730 & 1.2209 & 0.0018 \\
53\,862.6200 & 1.2217 & 0.0015 \\
53\,866.6037 & 1.2239 & 0.0014 \\
53\,871.6270 & 1.2357 & 0.0015 \\
53\,883.5589 & 1.2397 & 0.0014 \\
53\,918.4979 & 1.2589 & 0.0017 \\
53\,920.5110 & 1.2613 & 0.0023 \\
\hline
\end{tabular}
\end{table}

The main characteristics of {\footnotesize HD}\,100777, {\footnotesize HD}\,190647, and 
{\footnotesize HD}\,221287 are listed in Table~\ref{tabstars}. Spectral types, apparent magnitudes $m_{\rm V}$, colour 
indexes $B-V$, astrometric parallaxes $\pi$, and distances $d$ are from the {\footnotesize HIPPARCOS} Catalogue \citep{ESA97}. 
From the same source, we have also retrieved information on the scatter in the photometric measurements and on the goodness of 
the astrometric fits for the 3 targets. The photometric scatters are low in all cases. {\footnotesize HD}\,190647 is 
flagged as a constant star. The goodness-of-fit parameters are close to 0 for the 3 stars, indicating that their astrometric data 
are explained by a single-star model well. Finally, no close-in faint visual companions are reported around these objects in the 
{\footnotesize HIPPARCOS} Catalogue.

We performed  LTE spectroscopic analyses of high signal-to-noise ratio ({\footnotesize SNR}) {\footnotesize HARPS} spectra for 
the 3 targets fol\-low\-ing the method described in \citet{Santosmet3}. Effective temperatures ($T_{\rm eff}$), gravities ($\log g$), 
iron abundances ($[$Fe/H$]$), and microturbulence velocities ($V_{\rm t}$) indicated in Table~\ref{tabstars} result from these 
analyses. Like most of the planet-hosting stars \citep{Santosmet3}, {\footnotesize HD}\,100777 and 
{\footnotesize HD}\,190647 have very high iron abundances, more than 1.5 times the solar value, whereas 
{\footnotesize HD}\,221287 has a nearly solar metal content. Using the spectroscopic effective temperatures and the 
calibration in \citet{Flower96}, we computed bolometric corrections. Luminosities were obtained from the bolometric corrections 
and the absolute ma\-gni\-tu\-des. The low gravity and the high luminosity of {\footnotesize HD}\,190647 indicate that this 
star is slightly evolved. The other two stars are still on the main sequence. Stellar masses $M_*$ were derived from $L$, 
$T_{\rm eff}$, and $[$Fe/H$]$ using Geneva and Padova evo\-lu\-tiona\-ry models \citep{Schaller92,Schaerer93,Girardi2000}. Values of the 
projected rotational velocities, $v\sin i$, were derived from the widths of the {\footnotesize HARPS} cross-correlation functions 
using a calibration obtained following the method de\-scribed in 
\citet[][ see their Appendix~A]{Santoscalibvsini}~\footnote{Using cross-correlation function widths for deriving projected 
rotational velocities is a method that was first proposed by \citet{Benz81}.}

Stellar activity indexes $\log R^{'}_{HK}$ \citep[see the index definition in][]{Noyes84} were derived from \ion{Ca}{II}\,K line 
core re-emission measurements on high-{\footnotesize SNR} {\footnotesize HARPS} spectra following the method described in 
\citet{Santosact}. Using these va\-lues and the calibration in \citet{Noyes84}, we derived estimates of the rotational periods and 
stellar ages. The chromospheric ages obtained with this ca\-libration for {\footnotesize HD}\,100777 and 
{\footnotesize HD}\,190647 are 6.2 and 7.6\,Gyr. \citet{Pace2004} have shown that chromospheric ages derived for very 
low-activity, Solar-type stars were not reliable. This is due to the fact that the chromospheric emission drops rapidly after 
1\,Gyr and becomes virtually constant after 2\,Gyr. For this reason, we have chosen to indicate ages greater than 2\,Gyr instead 
of the calibrated values for these two stars. {\footnotesize HD}\,221287 is much more active and thus younger making its 
chromospheric age estimate more reliable: 1.3\,Gyr. We have also measured the lithium abundance for this target following the 
method described in \citet{Israelian2004} and again using a high-{\footnotesize SNR HARPS} spectrum. The measured equivalent width 
for the \ion{Li}{I} $\lambda$6707.70$\AA$ line (deblended from the \ion{Fe}{I} $\lambda$6707.44$\AA$ line) is 66.6\,m$\AA$ leading 
to the following lithium abundance: $\log n{(\rm Li)}$\,=\,2.98. Unfortunately, the lithium abundance cannot provide any reliable 
age constraint in our case. Studies of the lithium abundances of open cluster main sequence stars have shown that 
$\log n{\rm (Li)}$ remains constant and equals $\simeq 3$ for $T_{\rm eff}$\,=\,6300\,K stars older than a few million years 
\citep[see for example][]{Sestito2005}.

From the activity level reported for {\footnotesize HD}\,221287 ($\log R^{'}_{HK}$\,=\,$-$4.59), we can estimate the 
expected level of activity-induced radial-velocity scatter (i.e. the {\sl jitter}). Using the results obtained by 
\citet{Santosact} for stars with similar activity levels and spectral types, the range of expected jitter is between 8 and 
20\,m\,s$^{\rm -1}$. A similar study made on a different stellar sample by \citet{Wright2005} gives similar results: an expected 
jitter of the order of 20\,m\,s$^{\rm -1}$ (from the fit this author presents in  his Fig.~4). It has to be noted that both 
studies contain very few F stars and even fewer high-activity F stars. This results from the stellar sample they have used: 
planet search samples selected against rapidly-rotating young stars. Their predictions for the expected jitter level are therefore 
not well-constrained for active F stars. Both {\footnotesize HD}\,100777 and {\footnotesize HD}\,190647 are 
inactive and slowly rotating. Following the same studies, low jitter values are expected in both cases: 
$\leq$\,8\,m\,s$^{\rm -1}$.

\section{HARPS radial-velocity data}\label{RVdata}

\begin{figure}[t!]
   \centering
   \resizebox{0.95\hsize}{!}{\includegraphics{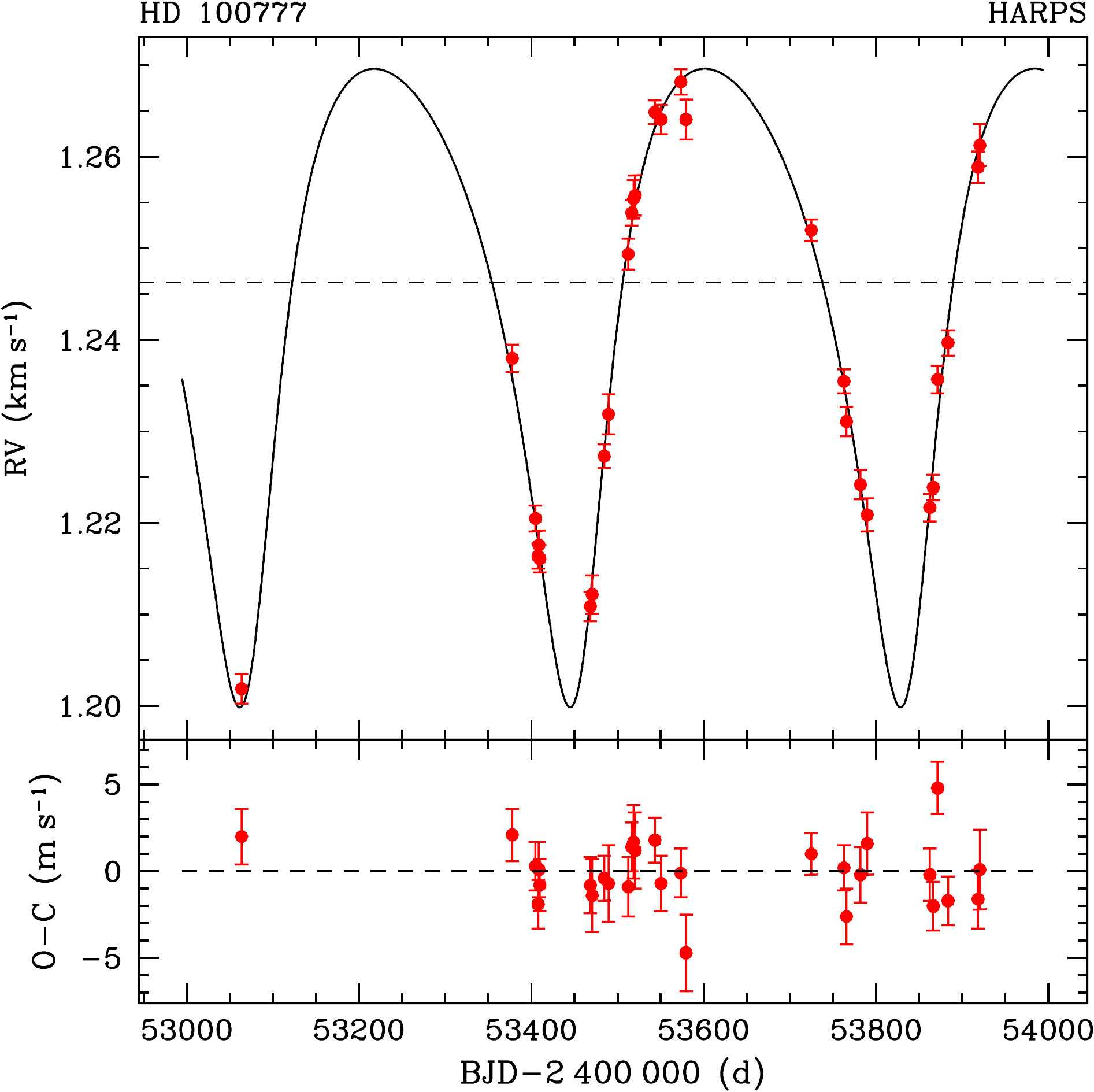}}
   \caption{{\bf Top:} {\footnotesize HARPS} radial-velocity data (dots) for {\footnotesize HD}\,100777 
   and fitted orbital solution (solid curve). The radial-velocity signal is induced by the presence of a
   1.16\,M$_{\rm Jup}$ planetary companion on a 384-day orbit. {\bf Bottom:} Residuals to the fitted 
   orbit. The scatter of these residuals is compatible with the velocity uncertainties.}
   \label{orb_hd100777}
\end{figure}

\begin{table}[t!]
\caption{
\label{hd190647data}
HARPS radial-velocity data obtained for HD\,190647.
}
\begin{tabular}{ccc}
\hline\hline
\noalign{\vspace{0.05cm}}
Julian date & RV & Uncertainty \\
BJD\,$-$\,2\,400\,000 $[$d$]$ & \multicolumn{2}{c}{$[$km\,s$^{\rm -1}]$}\\
\hline
\noalign{\vspace{0.05cm}}
\noalign{\vspace{0.05cm}}
52\,852.6233 & $-$40.2874 & 0.0013 \\
52\,854.6727 & $-$40.2868 & 0.0013 \\
53\,273.5925 & $-$40.2435 & 0.0022 \\
53\,274.6041 & $-$40.2354 & 0.0024 \\
53\,468.8874 & $-$40.2688 & 0.0015 \\
53\,470.8634 & $-$40.2689 & 0.0014 \\
53\,493.9234 & $-$40.2700 & 0.0029 \\
53\,511.9022 & $-$40.2723 & 0.0015 \\
53\,543.8020 & $-$40.2785 & 0.0013 \\
53\,550.7721 & $-$40.2804 & 0.0014 \\
53\,572.8074 & $-$40.2818 & 0.0015 \\
53\,575.7266 & $-$40.2844 & 0.0017 \\
53\,694.5136 & $-$40.3056 & 0.0014 \\
53\,836.9226 & $-$40.2982 & 0.0015 \\
53\,861.8630 & $-$40.2950 & 0.0017 \\
53\,883.8224 & $-$40.2907 & 0.0013 \\
53\,886.8881 & $-$40.2883 & 0.0013 \\
53\,917.8390 & $-$40.2785 & 0.0021 \\
53\,921.8472 & $-$40.2785 & 0.0018 \\
53\,976.6039 & $-$40.2623 & 0.0014 \\
53\,979.6654 & $-$40.2610 & 0.0017 \\
\hline
\end{tabular}
\end{table}

Stars belonging to the volume-limited {\footnotesize HARPS-GTO} sub-programme are observed most of the time without the 
simultaneous Thorium-Argon reference \citep{Baranne96}. The obtained radial velocities are thus uncorrected for possible 
ins\-tru\-men\-tal drifts. This only has a very low impact on our results as the {\footnotesize HARPS} radial-velocity drift is less 
than 1\,m\,s$^{\rm -1}$ over one night. For this large volume-limited sample, we aim at a radial-velocity precision of the order 
of 3\,m\,s$^{\rm -1}$ (or better). This cor\-res\-ponds roughly to an {\footnotesize SNR} of 40-50 at 5500$\AA$. For bright 
targets like the three stars of this paper, the exposure times required for reaching this signal level can be very short as an 
{\footnotesize SNR} of 100 (at 5500$\AA$) is obtained with {\footnotesize HARPS} in a 1-minute exposure on a 6.5\,mag G dwarf 
under normal weather and seeing conditions. In order to limit the impact of observing overheads (telescope preset, target 
acquisition, detector read-out), we normally do not use exposure times less than 60 seconds. As a consequence, the 
{\footnotesize SNR} obtained for bright stars is significantly higher than the targeted one, and the output measurement 
errors are frequently below 2\,m\,s$^{\rm -1}$. 

For our radial-velocity measurements, we consider two main error terms. The first one is obtained through the 
{\footnotesize HARPS} Data Reduction Software ({\footnotesize DRS}). It includes all the known calibration errors 
($\simeq$\,20\,cm\,s$^{\rm -1}$), the stellar photon-noise, and the error on the instrument drift. For observations taken with the 
simultaneous reference, the drift error is derived from the photon noise of the Thorium-Argon exposure. For observations taken 
without the lamp, a drift error term of 50\,cm\,s$^{\rm -1}$ is quadratically added. The second main error term is called the 
non-photonic error. It includes guiding errors and a lower limit for the stellar pulsation signals. For the volume-limited 
programme, we use an ad-hoc value for this term: 1.0\,m\,s$^{\rm -1}$. Stellar noise (activity jitter, pulsation signal) can of 
course be greater in some cases (as for example for {\footnotesize HD}\,221287, cf. Sect.~\ref{HD221287orbsol}). The 
non-photonic term nearly vanishes for targets belonging to the very high {\footnotesize RV} precision sample (non-active stars, 
pulsation modes averaged out by the specific observing stra\-te\-gy). In this latter case, this term thus only contains the 
guiding errors: $\simeq$\,30\,cm\,s$^{\rm -1}$. The error bars listed in this paper correspond to the quadratic sum of the 
{\footnotesize DRS} and the non-photonic errors.

In the following sections, we present the {\footnotesize HARPS} radial-velocity data obtained for 
{\footnotesize HD}\,100777, {\footnotesize HD}\,190647, and {\footnotesize HD}\,221287 in more detail, 
as well as the orbital solutions fitted to the data. 


\begin{figure}[t!]
   \centering
   \resizebox{0.95\hsize}{!}{\includegraphics{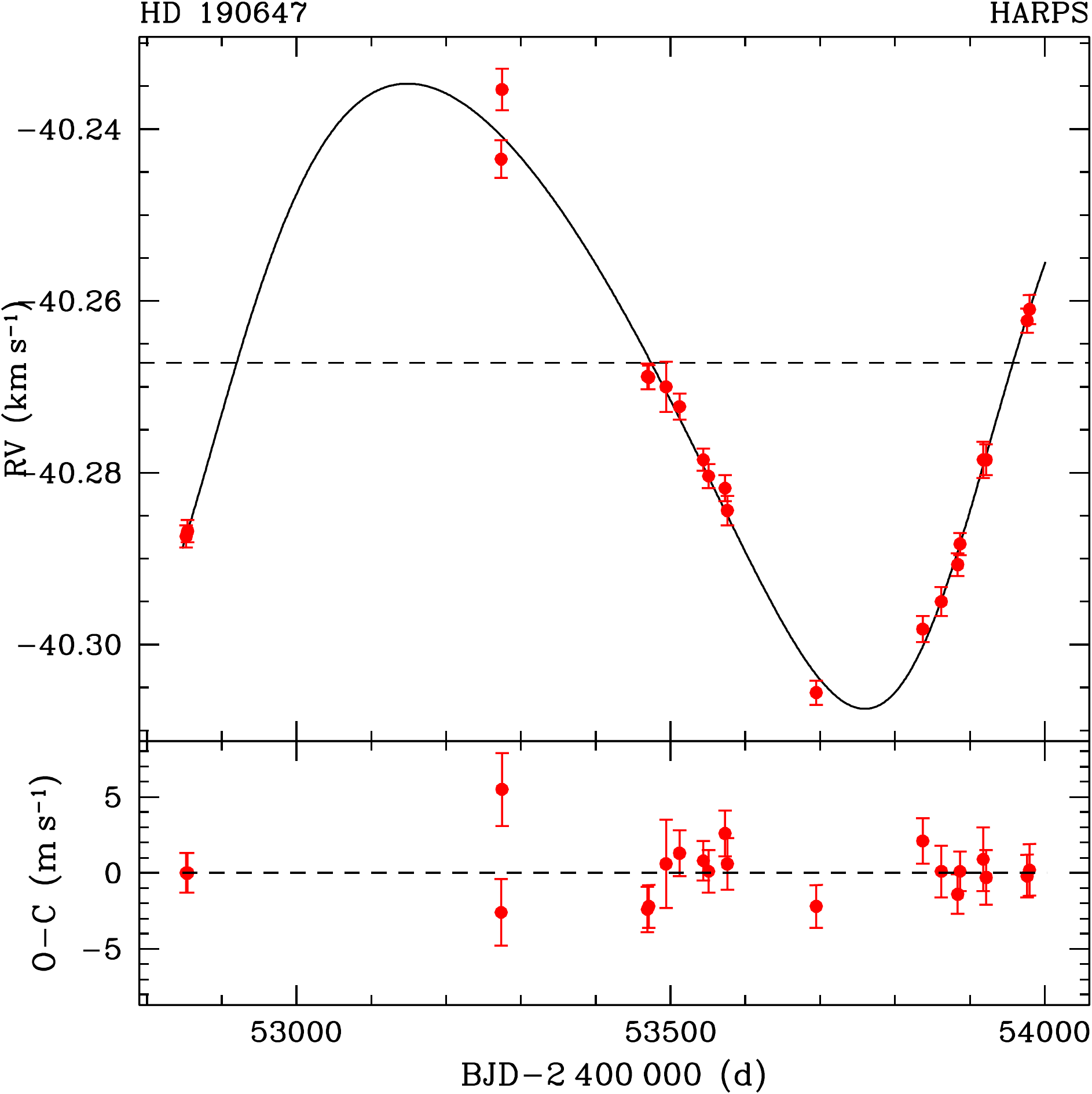}}
   \caption{{\bf Top:} {\footnotesize HARPS} radial-velocity data (dots) for {\footnotesize HD}\,190647 
   and fitted orbital solution (solid curve). The radial-velocity signal is induced by the presence of a
   1.9\,M$_{\rm Jup}$ planetary companion on a 1038-day orbit. {\bf Bottom:} Residuals to the fitted 
   orbit. The scatter of these residuals is compatible with the velocity uncertainties.}
   \label{orb_hd190647}
\end{figure}

\subsection{A 1.16\,M$_{\rm Jup}$ planet around HD\,100777}\label{HD100777orbsol}

We have gathered 29 {\footnotesize HARPS} radial-velocity measurements of {\footnotesize HD}\,100777. These data span 
858 days between February 27th 2004 ({\footnotesize BJD}\,=\,2\,453\,063) and July 4th 2006 ({\footnotesize BJD}\,=\,2\,453\,921). 
Their mean radial-velocity uncertainty is 1.6\,m\,s$^{\rm -1}$ (mean {\footnotesize DRS} error: 1.3\,m\,s$^{\rm -1}$). We list 
these measurements in Table~\ref{hd100777data} (electronic version only).

A nearly yearly signal is present in these data. We fitted a Keplerian orbit. The resulting parameters are listed  in 
Table~\ref{taborbsol}. The fitted orbit is displayed in Fig.~\ref{orb_hd100777}, together with our radial-velocity measurements. 
The radial-velocity data is best explained by the presence of a  1.16\,M$_{\rm Jup}$ planet on a 384\,d fairly eccentric orbit 
($e$\,=\,0.36). The inferred separation between the host star and its planet is $a$\,=\,1.03\,AU. Both $m_{\rm 2}\sin i$ and 
$a$ were computed using a primary mass of 1\,M$_{\sun}$.

We performed Monte-Carlo simulations using the {\footnotesize {\sl ORBIT}} software \citep[see Sect.~3.1. in ][]{Forveille1999} in 
order to double-check the parameter uncertainties. The errorbars obtained from these simulations are quasi-symmetric and somewhat 
larger ($\simeq$18\%) than the ones obtained from the covariance matrix of the  Keplerian fit. The errors we have finally quoted 
in Table~\ref{taborbsol} are the Monte-Carlo ones. The residuals to the fitted orbit (see bottom panel of Fig.~\ref{orb_hd100777}) 
are flat and have a dispersion compatible with the measurement noise. The low reduced $\chi^{\rm 2}$ value (1.45) and the 
$\chi^{\rm 2}$ probability (0.074) further demonstrate the good fit quality. The presence of another massive short-period 
companion around {\footnotesize HD}\,100777 is thus unlikely.

\begin{table}[t!]
\caption{
\label{hd221287data}
HARPS radial-velocity data obtained for HD\,221287.
}
\begin{tabular}{ccc}
\hline\hline
\noalign{\vspace{0.05cm}}
Julian date & RV & Uncertainty \\
BJD\,$-$\,2\,400\,000 $[$d$]$ & \multicolumn{2}{c}{$[$km\,s$^{\rm -1}]$}\\
\hline
\noalign{\vspace{0.05cm}}
52\,851.8534 & $-$21.9101 & 0.0012 \\
52\,853.8544 & $-$21.9201 & 0.0021 \\
52\,858.7810 & $-$21.9252 & 0.0019 \\
53\,264.7097 & $-$21.8617 & 0.0021 \\
53\,266.6805 & $-$21.8852 & 0.0020 \\
53\,268.7030 & $-$21.8595 & 0.0034 \\
53\,273.6951 & $-$21.8888 & 0.0031 \\
53\,274.7129 & $-$21.8718 & 0.0021 \\
53\,292.6284 & $-$21.9049 & 0.0020 \\
53\,294.6239 & $-$21.8998 & 0.0019 \\
53\,295.6781 & $-$21.9017 & 0.0024 \\
53\,296.6388 & $-$21.9186 & 0.0026 \\
53\,339.6009 & $-$21.9289 & 0.0015 \\
53\,340.5974 & $-$21.9225 & 0.0013 \\
53\,342.5955 & $-$21.9172 & 0.0013 \\
53\,344.5961 & $-$21.9428 & 0.0014 \\
53\,345.5923 & $-$21.9291 & 0.0013 \\
53\,346.5566 & $-$21.9284 & 0.0013 \\
53\,546.9385 & $-$21.8022 & 0.0039 \\
53\,550.9121 & $-$21.8294 & 0.0022 \\
53\,551.9479 & $-$21.7956 & 0.0018 \\
53\,723.5707 & $-$21.8679 & 0.0033 \\
53\,727.5302 & $-$21.8815 & 0.0021 \\
53\,862.9297 & $-$21.9205 & 0.0023 \\
53\,974.7327 & $-$21.8240 & 0.0021 \\
53\,980.7273 & $-$21.8192 & 0.0022 \\
\noalign{\vspace{0.05cm}}
\hline
\end{tabular}
\end{table}
\subsection{A 1.9\,M$_{\rm Jup}$ planet orbiting HD\,190647}\label{HD190647orbsol}

Between August 1, 2003 ({\footnotesize BJD}\,=\,2\,452\,852) and September 2, 2006 ({\footnotesize BJD}\,=\,2\,453\,980), we  
obtained 21 {\footnotesize HARPS} radial-velocity measurements for {\footnotesize HD}\,190647. These data have a mean 
radial-velocity uncertainty of 1.7\,m\,s$^{\rm -1}$ (mean {\footnotesize DRS} error: 1.3\,m\,s$^{\rm -1}$). We list these 
measurements in Table~\ref{hd190647data} (electronic version only).

A long-period signal is clearly present in the {\footnotesize RV} data. We performed a Keplerian fit. The resulting fitted 
parameters are listed in Table~\ref{taborbsol}. The fitted orbital period (1038\,d) is slightly shorter than the observing time 
span (1128\,d) and the orbital eccentricity is low (0.18). Monte-Carlo si\-mu\-la\-tions were carried out. The uncertainties on 
the orbital parameters obtained in this case are nearly symmetric and a bit larger ($\simeq$18\%) than the ones resulting from the 
Keplerian fit. In Table~\ref{taborbsol}, we have listed these more conservative Monte-Carlo uncertainties. The small discrepancy 
between the two sets of errorbars is most probably due to the rather short time span of the observations (only 1.09 orbital cycle 
covered) and to our still not optimal coverage of both the minimum and the maximum of the radial-velocity orbit. From the fitted 
parameters and with a primary mass of 1.1\,M$_{\sun}$, we compute a minimum mass of 1.90\,M$_{\rm Jup}$ for this planetary 
companion. The computed separation between the two bodies is 2.07\,AU. Figure~\ref{orb_hd190647} shows our data and the fitted 
orbit.

The weighted {\sl rms} of the residuals (1.6\,m\,s$^{\rm -1}$) is slightly smaller than the mean {\footnotesize RV} uncertainty. 
The low dispersion of the residuals, the low reduced $\chi^2$ value, and its associated probability (0.11) allow us to exclude the 
presence of an additional massive short-period companion.

\subsection{A 3.1\,M$_{\rm Jup}$ planetary companion to HD\,221287}\label{HD221287orbsol}

\begin{figure}[t!]
   \centering
   \resizebox{0.95\hsize}{!}{\includegraphics{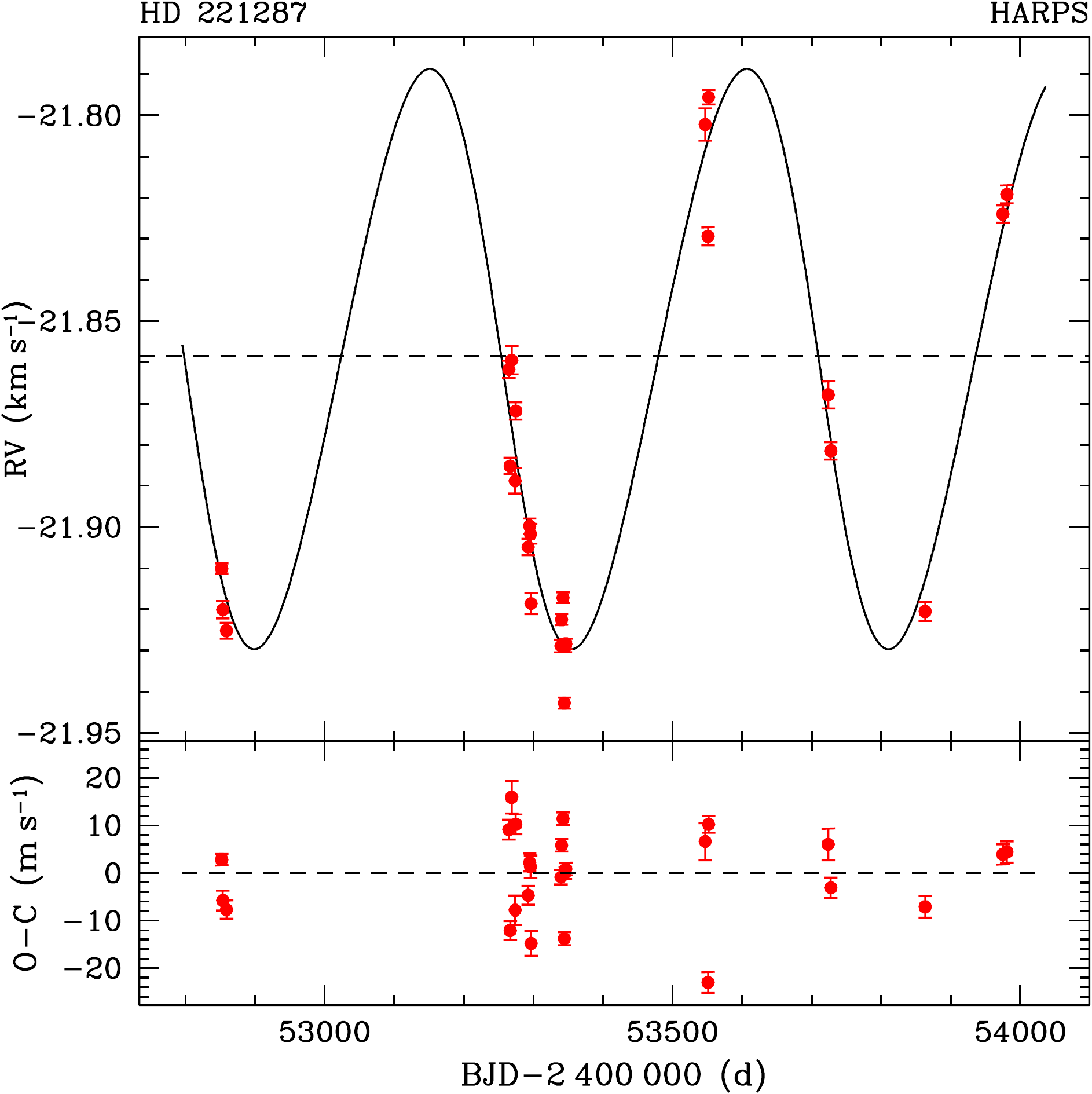}}
   \caption{{\bf Top:} {\footnotesize HARPS} radial-velocity data (dots) for {\footnotesize HD}\,221287 
   and fitted orbital solution (solid curve). The detected Keplerian signal is induced by a 3.1\,M$_{\rm Jup}$ 
   planet on a 456--day orbit. 
   Because of a non-optimal coverage of the radial-velocity maximum
   and the presence of stellar activity-induced jitter, the exact shape of the orbit is not very well-constrained. 
   {\bf Bottom:} Residuals to the fitted orbit. The scatter of these residuals is much larger than the velocity 
   uncertainties. This large dispersion is probably due to the fairly high activity level of this star.}
   \label{orb_hd221287}
\end{figure}

A total of 26 {\footnotesize HARPS} radial-velocity data were obtained for {\footnotesize HD}\,221287. These data are 
spread over 1130 days: between July 31, 2003 ({\footnotesize BJD}\,=\,2\,452\,851) and September 3, 2006 
({\footnotesize BJD}\,=\,2\,453\,981). Unlike the two other targets presented in this paper, a substantial fraction ($\simeq$65\%) 
of these velocities were taken using the simultaneous Thorium-Argon reference. They were thus corrected for the measured 
instrumental velocity drifts. The mean radial-velocity uncertainty computed for this data set is 2.1\,m\,s$^{\rm -1}$ (mean 
{\footnotesize DRS} error: 1.8\,m\,s$^{\rm -1}$). We list these measurements  in Table~\ref{hd221287data} (electronic version 
only).

\begin{table*}[t!]
\caption{
\label{taborbsol}
HARPS orbital solutions for HD\,100777, HD\,190647, and HD\,221287.
}
\begin{tabular}{l|c|r@{\,$\pm$\,}l|r@{\,$\pm$\,}l|r@{\,$\pm$\,}l}
\hline
\hline
\noalign{\vspace{0.05cm}}
                                   &                            &\multicolumn{2}{c|}{HD\,100777} & \multicolumn{2}{c|}{HD\,190647} & \multicolumn{2}{c}{HD\,221287} \\
\hline
\noalign{\vspace{0.05cm}}
$P$                                &  $[$d$]$                       & 383.7         & 1.2            & 1038.1      & 4.9                & 456.1       & $^{+{\rm 7.7}}_{-{\rm 5.8}}$\\[0.1cm]
$T$                                &  $[$JD$^{\dagger}$$]$          & 456.2         & 2.3            & 868         & 24                 & 263         & $^{+{\rm 99}}_{-{\rm 52}}$\\[0.1cm]
$e$                                &                                & 0.36          & 0.02           & 0.18        & 0.02               & 0.08        & $^{+{\rm 0.17}}_{-{\rm 0.05}}$ \\[0.1cm]
$\gamma$                           &  $[$km\,s$^{\rm -1}$$]$        & 1.246         & 0.001          & $-$40.267   & 0.001              & $-$21.858   & $^{+{\rm 0.008}}_{-{\rm 0.005}}$\\[0.1cm]
$\omega$                           &  $[$$\degr$$]$                 & 202.7         & 3.1            & 232.5       & 9.4                & 98          & $^{+{\rm 92}}_{-{\rm 51}}$\\[0.1cm]
$K_{\rm 1}$                        &  $[$m\,s$^{\rm -1}$$]$         & 34.9          & 0.8            & 36.4        & 1.2                & 71          & $^{+{\rm 18}}_{-{\rm 8}}$\\[0.1cm]
$f(m)$                             &  $[$10$^{\rm -9}$M$_{\sun}$$]$ & 1.37          & 0.10           & 4.94        & 0.50               & 16.4        & 12.6\\
$a_{\rm 1}\sin i$                  &  $[$10$^{\rm -3}$AU$]$         & 1.15          & 0.03           & 3.42        & 0.12               & 2.95        & 0.75\\[0.1cm]
\noalign{\vspace{0.05cm}}
\hline
\noalign{\vspace{0.05cm}}
$m_{\rm 2}\sin i$                  &  $[$M$_{\rm Jup}$$]$           & 1.16          & 0.03           & 1.90        & 0.06               & 3.09        & 0.79\\[0.1cm]
$a$                                &  $[$AU$]$                      & 1.03          & 0.03           & 2.07        & 0.06               & 1.25        & 0.04\\
\noalign{\vspace{0.05cm}}
\hline
\noalign{\vspace{0.05cm}}
$N$                                &                                & \multicolumn{2}{c|}{29}        & \multicolumn{2}{c|}{21}         & \multicolumn{2}{c}{26}\\[0.1cm]
$\sigma_{\rm O-C}$$^{\natural}$    &  $[$m\,s$^{\rm -1}$$]$         & \multicolumn{2}{c|}{1.7}       & \multicolumn{2}{c|}{1.6}        & \multicolumn{2}{c}{8.5}\\[0.1cm]
$\chi^{\rm 2}_{\rm red}$$^{\star}$ &                                & \multicolumn{2}{c|}{1.45}      & \multicolumn{2}{c|}{1.46}       & \multicolumn{2}{c}{26.7}\\[0.1cm]
$p(\chi^{\rm 2},\nu)$$^{\ddagger}$      &                         & \multicolumn{2}{c|}{0.074}     & \multicolumn{2}{c|}{0.11}       & \multicolumn{2}{c}{0}\\
\noalign{\vspace{0.05cm}}
\hline
\noalign{\vspace{0.05cm}}
\end{tabular}

$^\dagger$ JD\,=\,BJD\,$-$\,2\,453\,000\\[0.05cm]
$^\natural$ $\sigma_{\rm O-C}$ is the weighted {\sl rms}   of the residuals (weighted by $1/\epsilon^{\rm 2}$, where $\epsilon$ is the O$-$C uncertainty)\\[0.05cm]
$^\star$ $\chi^{\rm 2}_{\rm red}$\,=\,$\chi^{\rm 2}/\nu$ where $\nu$ is the number of degrees of freedom (here $\nu$\,=\,$N-6$).\\[0.05cm]
$^\ddagger$ Post-fit $\chi^{\rm 2}$ probability computed with $\nu$\,=\,$N-6$.
\end{table*}

A 456\,d radial-velocity variation is clearly visible in our data (see Fig.~\ref{orb_hd221287}). This period is two orders of 
magnitude longer than the rotation period obtained from the \citet{Noyes84} calibration for {\footnotesize HD}\,221287: 
5\,$\pm$\,2\,d. This large discrepancy between $P_{\rm RV}$ and $P_{\rm rot}$  is probably  sufficient for safely excluding 
stellar spots as the origin of the detected {\footnotesize RV} signal, but  we nevertheless checked if this variability could be 
due to line-profile variations. The cross-correlation function ({\footnotesize CCF}) bisector span versus radial velocity 
plot is shown in the top panel of Fig.~\ref{bisec_hd221287}.  The average {\footnotesize CCF} bisector value is computed in two 
selected regions: near the top of the {\footnotesize CCF} (i.e. near the continuum) and near its bottom (i.e near the 
{\footnotesize RV} minimum). The span is the difference between these two average values (top$-$bottom) and thus represents the 
overall slope of the {\footnotesize CCF} bisector \citep[for details, see][]{Quelozhd166435}.

As for the case of {\footnotesize HD}\,166435 presented in \citet{Quelozhd166435}, an anti-correlation between spans and 
velocities is expected in the case of star-spot induced line-profile variations. The bisector span data are quite noisy (weighted 
{\sl rms} of 10.4\,m\,s$^{\rm -1}$), but they are not correlated with the {\footnotesize RV} data. The main signal can therefore 
not be due to line-profile variations and certainly has a Keplerian origin.

Table~\ref{taborbsol} contains the results of a Keplerian fit that we performed. Our data and the fitted orbit are displayed in 
Fig.~\ref{orb_hd221287}. The {\footnotesize RV} maximum remains poorly covered by our observations. As for the other two targets, 
we made Monte-Carlo simulations for checking our orbital parameter uncertainties. As expected, the uncertainties obtained in this 
case largely differ from the ones obtained via the Keplerian fit. For most of the parameters, the errorbars resulting from the 
simulations are not symmetric and much larger ($\simeq$5 times larger). In order to be more conservative, we have chosen to quote 
these errors in Table~\ref{taborbsol}. The shape of the orbit is not very well-constrained, but there is no doubt about the 
planetary nature of {\footnotesize HD}\,221287\,b. The fitted eccentricity is low (0.08), but circular or moderately 
eccentric orbits (up to 0.25) cannot be excluded yet. Using a primary mass of 1.25\,M$_{\sun}$, we compute the companion minimum 
mass and separation: $m_{\rm 2}\sin i$\,=\,3.09\,M$_{\rm Jup}$ and $a$\,=\,1.25\,AU.

\subsection{Residuals to the HD\,221287 orbital fit}\label{oc221287}

\begin{figure}[t!]
   \centering
   \resizebox{0.9\hsize}{!}{\includegraphics{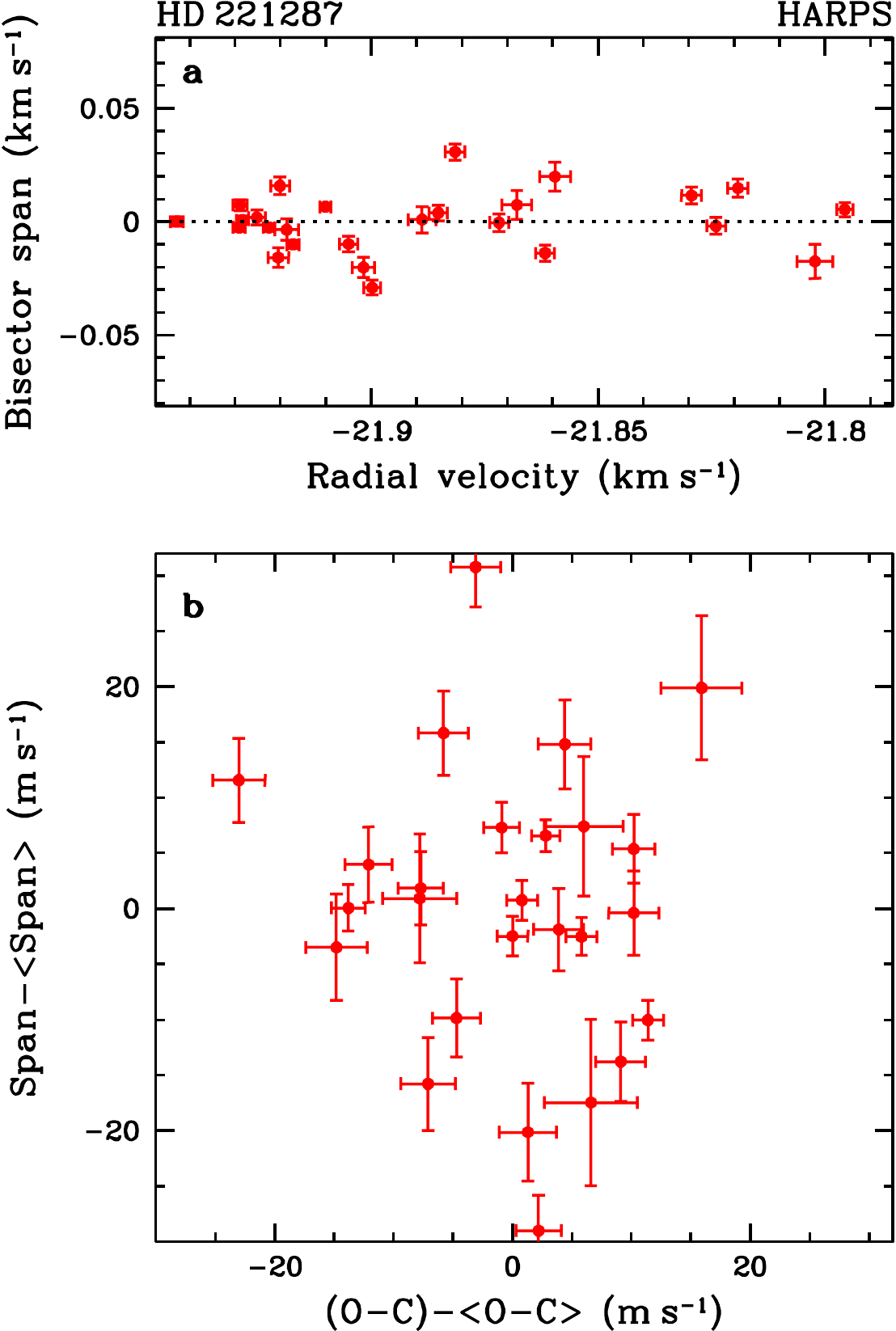}}

   \caption{{\bf a:} Bisector span versus radial-velocity plot for {\footnotesize HD}\,221287. 
   The dispersion of the span data is quite large (10.4\,m\,s$^{\rm -1}$) revealing potential line-profile
   variations. Nevertheless, the main radial-velocity signal is not correlated to these profile
   variations and is thus certainly of Keplerian origin. {\bf b:} Bisector span versus radial-velocity
   residuals to the Keplerian orbit (see Table~\ref{taborbsol}) displayed in the same velocity scale. 
   A marginal anti-correlation between the two quantities is observed. 
   }
   \label{bisec_hd221287}
\end{figure}

The residuals to the orbital solution for {\footnotesize HD}\,221287 presented in Sect.~\ref{HD221287orbsol} are clearly 
abnormal. Their weighted {\sl rms}, 8.5\,m\,s$^{\rm -1}$, is much larger than the mean radial-velocity uncertainty obtained for 
this target: $\langle\epsilon_{\rm RV}\rangle$\,=\,2.1\,m\,s$^{\rm -1}$.  The abnormal scatter obtained by quadratically 
correcting the residual {\sl rms}  for $\langle\epsilon_{\rm RV}\rangle$ is 8.2\,m\,s$^{\rm -1}$. This matches the lowest value 
expected for this star from the \cite{Santosact} and \citet{Wright2005} studies. Again, we stress that these two studies clearly 
lack active F stars, and their activity versus jitter relations are thus weakly constrained for this kind of target. Our measured 
jitter value certainly does not strongly disagree with their results.

We have searched for periodic signals in the radial-velocity residuals by computing their Fourier transform, but no significant 
peak in the power spectrum could be found. The absence of significant periodicity is not surprising since the phase of star-spot 
induced signals is not always conserved over more than a few rotational cycles. 

Cross-correlation function bisector spans are plotted against the observed radial-velocity residuals in the bottom panel of 
Fig.~\ref{bisec_hd221287}. A marginal anti-correlation (Spearman's rank correlation coefficient: $\rho$\,=\,$-$0.1) between the 
two quantities is visible.  A weighted linear regression (i.e the simplest  possible model) was computed. The obtained slope is 
only weakly significant (1$\sigma$). We are thus unable, at this stage, to clearly establish the link between the line-profile 
variations and our residuals. As indicated in Sect.~\ref{HD221287orbsol}, our orbital solution is not very well-constrained. 
This probably affects the residuals and possibly explains the absence of a clear anti-correlation. Additional radial-velocity 
measurements are necessary for establishing this relation, but activity-related processes so far remain the best explanation for 
the observed abnormal residuals to the fitted orbit.

{\footnotesize HD}\,221287 has a planet with an orbital period of 456\,d but with an additional radial-velocity signal, 
probably induced by the presence of cool spots whose visibility is modulated by stellar rotation.

\section{Conclusion}\label{Conc}

We have presented our {\footnotesize HARPS} radial-velocity data for 3 Solar-type stars: {\footnotesize HD}\,100777, 
{\footnotesize HD}\,190647, and {\footnotesize HD}\,221287. The radial-velocity variations detected for these 
stars are explained by the presence of planetary companions. {\footnotesize HD}\,100777\,b has a minimum mass of 
1.16\,M$_{\rm Jup}$. Its orbit is eccentric (0.36) and has a period of 384 days. The 1038--day orbit of the 1.9\,M$_{\rm Jup}$ 
planet around the slightly evolved star {\footnotesize HD}\,190647 is moderately eccentric (0.18). The planetary 
companion inducing the detected velocity signal for {\footnotesize HD}\,221287 has a minimum mass of 3.1\,M$_{\rm Jup}$. 
Its orbit has a period of 456 days. The orbital eccentricity for this planet is not well-constrained. The fitted value is 0.08 but 
orbits with 0.0\,$\leq$\,$e$\,$\leq$\,0.25 cannot be excluded yet. This rather weak constraint on the orbital shape is explained 
by two reasons. First, our data cover the radial-velocity maximum poorly. Second, the residuals to this orbit are abnormally 
large. We have tried to establish the relation between these high residuals and line-profile variations through a study of the 
{\footnotesize CCF} bisectors. As expected, a marginal anti-correlation of the two quantities is observed, but it is only 
weakly significant, thereby preventing us from clearly establishing the link between them.

\begin{acknowledgements}
The authors would like to thank the {\footnotesize ESO}--La Silla Observatory Science Operations team for its 
efficient support during the observations and to all the {\footnotesize ESO} staff involved in the 
{\footnotesize HARPS} maintenance and technical support. 
Support from the Funda\c{c}\~{a}o para Ci\^{e}ncia e a Tecnologia (Portugal) to N.C.S. in the form of a 
scholarship (reference SFRH/BPD/8116/2002) and a grant (reference POCI/CTEAST/56453/2004) is 
gratefully acknowledged. Continuous support from the Swiss National Science Foundation is 
appreciatively acknowledged. 
This research has made use of the Simbad database operated at the {\footnotesize CDS}, Strasbourg, France.
\end{acknowledgements}

\bibliographystyle{aa}
\bibliography{dnaef2007.bib}
\end{document}